\documentclass[floats,aps,showpacs]{revtex4}

\usepackage[dvips]{graphicx}
\usepackage{amsmath}   
\usepackage{amsfonts}

\newcommand{\rhob}{\overline{\rho}}
\newcommand{\rhot}{\widetilde{\rho}}
\newcommand{\Nbar}{\overline{{\cal N}}}
\newcommand{\Nosc}{\widetilde{{\cal N}}}
\newcommand{\tmin}{\tau_{\rm min}}
\newcommand{\tH}{\tau_{\scriptscriptstyle {\rm H}}}
\newcommand{\htop}{h_{\scriptscriptstyle {\rm top}}}

\newcommand{\ue}{\text{e}}
\newcommand{\ui}{\text{i}}
\newcommand{\tr}{\text{tr}}
\newcommand{\imag}{\text{Im}}
\newcommand{\Ci}{\text{Ci}}

\begin{document}

\title{Generalized random matrix conjecture for chaotic systems}

\author{P. Leboeuf$^1$ and A. G. Monastra$^{2,3}$}

\affiliation{$^1$Laboratoire de Physique Th\'eorique et Mod\`eles Statistiques, CNRS, Universit\'e Paris Sud, UMR8626, 91405 Orsay Cedex, France\\ $^2$ Gerencia Investigaci\'on y Aplicaciones, Comisi\'on Nacional de Energ\'\i a At\'omica, Avda. General Paz 1499, (1650) San Mart\'\i n, Argentina\\  $^3$ Consejo Nacional de Investigaciones Cient\'\i ficas y T\'ecnicas, Avda. Rivadavia 1917, (1033) Buenos Aires, Argentina}


\begin{abstract}
The eigenvalues of quantum chaotic systems have been conjectured to follow, in the large energy limit, the statistical distribution
of eigenvalues of random ensembles of matrices of size $N\rightarrow\infty$. Here we provide semiclassical arguments that extend the validity of this correspondence to finite energies. We conjecture that the spectrum of a generic fully chaotic system without time-reversal symmetry has, around some large but finite energy $E$, the same statistical properties as the Circular Unitary Ensemble of random matrices of dimension $N_{\rm eff} = \tH / \sqrt{24 d_1}$, where $\tH$ is Heisenberg time and $\sqrt{d_1}$ is a characteristic classical time, both evaluated at energy $E$. A corresponding conjecture is also made for chaotic maps.
\end{abstract}

\maketitle

\section{Introduction}

The study of connections between Random Matrix Theory (RMT) and properties of chaotic systems has now a long history  \cite{Berry85, reviewRMT}. A central point is the Bohigas-Giannoni-Schmit conjecture \cite{BGS}. For systems without time reversal invariance, the conjecture states that in the asymptotic high-energy limit the fluctuation properties of eigenvalues of quantum chaotic systems coincide with those of the Gaussian unitary ensemble (GUE) of random matrices of dimensionality $N\to\infty$. In particular, the normalized pair correlation function of the asymptotic eigenvalues of a chaotic system is conjectured to be
\begin{equation} \label{r2gue}
R_2 (x) = 1 -  \left( \frac{\sin  \pi x}{\pi x}  \right)^2\;,
\end{equation}
where $x$ is the unfolded distance between eigenvalues (i.e. the mean spacing has been set to one).

Nowadays, RMT has a large number of applications  \cite{applic}, ranging from disordered systems in quantum mechanics,
to growth processes in statistical physics and zeros of Riemann Zeta function in number theory. It is often
used to obtain deep conjectures or to compute particular properties of physical systems which are usually beyond the reach of exact methods. The validity of these predictions is then tested via experiments or via numerical calculations.

One of the difficulties concerning such an approach is that often real systems are not in the asymptotic regime required by the conjecture, and it may be experimentally quite unnatural to force them into it. Another point is related to the numerical tests of the predictions, which due to their asymptotic nature often require difficult large scale computations.

A relevant question concerns therefore the statistical properties of the finite-energy eigenvalues of quantum chaotic systems, and its eventual connections to RMT. In the following we will argue that such a connection exists. More precisely, based on a semiclassical analysis of the two--point correlation function we conjecture that the spectrum of a generic fully chaotic dynamical system without time-reversal symmetry has, around some finite energy $E$, the same statistical properties of the eigenvalues of the Circular Unitary Ensemble (CUE) of random matrices of dimension
\begin{equation}
N_{\rm eff} = \frac{ \tH }{ \sqrt{24 d_1} } \ ,  \label{NeffDS}
\end{equation}
where
\begin{equation}
d_1 = \frac{1}{2} \lim_{\tau_1 \rightarrow \infty} \left[ \frac{ \tau_1^2 }{2} - \sum_{\tau_p < \tau_1} \frac{ \tau_p^2 }{ r_p^2 \left| \det (M_p - I) \right| }  \right] \ . \label{d1}
\end{equation}
In Eq.~(\ref{NeffDS}), $\tH = h \rhob (E)$ is the Heisenberg time, where $\rhob (E)$ is the smooth part of the quantum density of states. The factor $\sqrt{d_1}$ is a system-dependent classical time expressed in terms of a sum over all the classical periodic orbits $p$ of the system at energy $E$, $\tau_p$ is the period of the periodic orbit, and $M_p$ it's monodromy matrix ($I$ is the identity matrix). The sum is over all periodic orbits of period less than $\tau_1$, including the repetitions $r_p$ when orbit $p$ is non--primitive ($r_p=1$ for primitive periodic orbits). Although not explicitly indicated, $N_{\rm eff}$ should be interpreted as the closest integer approximation of the r.h.s.. In the semiclassical limit $E\to\infty$, $N_{\rm eff} \rightarrow \infty$, and we recover the asymptotic CUE result  (that coincides with the Gaussian Unitary Ensemble), e.g. we recover the Bohigas-Giannoni-Schmit conjecture.

A similar problem may be considered for another type of systems, when the classical chaotic dynamics is described by a canonical map. These systems have been quite useful as schematic models for studying the quantum-classical correspondence, since a chaotic dynamics is possible in one dimension. When their phase-space is compact, they also allow to avoid truncations of the quantum mechanical model, since their evolution operator is given by a unitary matrix of finite dimension $N$, where the limit $N\to\infty$ describes the semiclassical limit. Connections between chaotic maps and random matrix ensembles have been already established \cite{Smilansky} in this limit. Beyond their formal character, maps have also been quite useful to analyze real systems since, sometimes, they provide quite good approximations to their dynamics \cite{realmaps}. Well known examples of such maps are the standard map \cite{standmap}, the baker map \cite{bakermap}, or the kicked Harper map \cite{harpermap}.

We have also made a semiclassical analysis of the statistical properties of the eigenphases of quantum chaotic maps without time-reversal invariance. For models described by a finite unitary matrix $\hat U$ of size $N$, we conjecture that the spectrum of these chaotic maps have the same statistical properties of CUE random matrices of dimension
\begin{equation}
N_{\rm eff} = \frac{ N }{ \sqrt{1 + 24 c_1} } \ ,  \label{Neffmaps}
\end{equation}
where
\begin{equation}
c_1 = \frac{1}{2} \lim_{n^* \rightarrow \infty} \sum_{n=1}^{n^*} ( n - |a_n|^2 )    \label{d1maps}
\end{equation}
is a finite real number that depends on the system considered ($a_n = \tr \hat U^{n}$ are the traces of powers of the evolution operator $\hat U$).

The present work may be viewed as a continuation of the results obtained in Ref.\cite{BBLM}, where it was conjectured that zeros of the Riemann zeta function located around $z= 1/2 + \ui E$ have the same statistical properties as the eigenvalues of CUE random matrices of effective dimension
\begin{equation}\label{neff}
N_{\rm eff} = \frac{\log (E / 2 \pi) }{\sqrt{12 \beta} }
\end{equation}
where $\beta = 1.57314$, is a constant depending on sums over prime numbers (see the discussion in Section~\ref{Conclusions}).

The paper is organized as follows. In Section~\ref{R2function} general expressions are given for the density of states, its semiclassical approximation, and the two--point correlation function. For chaotic Hamiltonian systems, the method developed by Bogomolny and Keating \cite{BogKea96} is used to express the two--point correlation function in terms of a classical function, which is formally exact. A similar method for maps is also developed.

In Section~\ref{DeltaFunction} the classical functions obtained in Section~\ref{R2function} are expanded for small arguments, which corresponds to the semiclassical limit, keeping all orders.

In Section~\ref{R2RMT} we first consider the exact two--point correlation function of the Circular Unitary Ensemble at finite dimension $N$. Secondly, the expansion of the classical function from previous section is used to compute the two--point correlation function up to second power in Heisenberg time for chaotic Hamiltonian systems without time reversal symmetry. Analogously we obtain a result for maps up to order $N^2$. Comparing these results to those of CUE, we derive the effective dimension.

Finally, in Section~\ref{Conclusions} we check the results for the Riemann Zeta function, making some general concluding remarks.

\section{Semiclassical calculation of the two--point correlation function} \label{R2function}

In the semiclassical limit, the density of states
\begin{equation}
\rho (E) = \sum_j \delta (E-E_j) \,
\label{DofS}
\end{equation}
may be split into a smooth plus an oscillatory part $\rho = \rhob + \rhot$. The smooth part admits an expansion in powers of energy called the Weyl series, that depends on global properties of the system such as volume, surface and topology of the classical phase space, disregarding the integrable or chaotic nature of the dynamics. The oscillatory part may, in contrast, be written
\begin{equation}
\rhot (E) = 2 \sum_p A_p (E) \cos \left[ S_p (E)/\hbar+ \nu_p \right] \ ,
\label{Gutz}
\end{equation}
where as in Eq. (\ref{d1}) the sum is over all classical periodic orbits $p$ of the system at energy $E$, $S_p$ is the corresponding action, and $\nu_p$ is a phase called the Maslov index that depends on the number of turns of the stable and unstable varieties along the periodic orbits.

The value of the amplitude $A_p$ depends on the stability of the periodic orbit. If the orbit is unstable, the amplitude is given by the Gutzwiller trace formula \cite{Gutzwiller}
\begin{equation}
A_p = \frac{\tau_p}{2\pi\hbar \ r_p \sqrt{\left| \det (M_p - I) \right|}} \ .
\label{Apchaotic}
\end{equation}

Several  statistical spectral functions are of interest. One of the most basic ones, on which we will concentrate on, is the two--point correlation function
\begin{equation}
r_2(\epsilon_1, \epsilon_2) =  \langle \rho (E + \epsilon_1) \rho (E + \epsilon_2) \rangle \ . \label{r2def}
\end{equation}
The brackets indicate an energy average defined as
\begin{equation}
\langle f (E) \rangle = \frac{1}{\Delta E} \int_{E-\Delta E/2}^{E+\Delta E/2} f(E') \rm d E' \ ,
\label{averagedef}
\end{equation}
where the energy window $\Delta E$ is small compared to $E$ but it is however large enough to be statistically meaningful (usually $\Delta E \gg 1/\rhob$).

\subsection{Hamiltonian systems} \label{R2semHamiltonian}

Semiclassically, the $r_2$ function may be computed by introducing the previous semiclassical approximation into the definition (\ref{r2def}). The energy average of the double sum over periodic orbits splits into a diagonal and an off-diagonal part. By the Hannay-Ozorio de Almeida sum rule the diagonal part can be easily estimated. The main problem is to compute the off-diagonal term. Actions of different periodic orbits have correlations that are not yet fully understood. To overcome this problem we follow the pioneering work of Bogomolny and Keating \cite{BogKea96}, where it is argue that the two--point correlation function of a chaotic system tends to the random matrix prediction in the semiclassical limit. We sketch here some details of the calculation in order to compute next to leading order corrections, taking into account all the approximations.

We begin by defining for chaotic systems a truncated density of states and counting function
\begin{eqnarray}
\rhot_{\tau^*} (E) &=& \frac{1}{2 \pi \hbar} \sum_{\tau_p < \tau^*} \frac{\tau_p \ \ue^{\ui (S_p/\hbar + \nu_p)} }{r_p \left| \det (M_p - I) \right|^{1/2}} + \rm{c.c.} \ ,  \label{rhotrunc} \\
\Nosc_{\tau^*} (E) &=& \frac{1}{2 \pi \ui} \sum_{\tau_p < \tau^*} \frac{ \ue^{\ui (S_p/\hbar + \nu_p)} }{r_p \left| \det (M_p - I) \right|^{1/2}} + \rm{c.c.} \ ,  \label{Ntrunc}
\end{eqnarray}
If $\tau^*$ is of the order of Heisenberg time, and
\begin{equation}
{\cal N}_{\tau^*} (E) = \Nbar(E) + \Nosc_{\tau^*} (E) = n + 1/2
\end{equation}
then the energies $E = E_n (\tau^*)$ obtained by inverting the latter equation with $n=0, 1, 2, \ldots$ provide a semiclassical approximation of the exact spectrum. Once we have these approximate eigenvalues we define a bootstrapped density of states

\begin{equation}
D_{\tau^*} (E) = \sum_n \delta [ E - E_n (\tau^*) ] \label{bootrho}
\end{equation}
As $\tau^* \rightarrow \infty$, it is obvious that $D_{\tau^*} (E) \rightarrow \rho (E)$. Rewriting the sum and using Poisson summation formula

\begin{equation}
D_{\tau^*} (E) = \rho_{\tau^*} (E) \sum_{k = -\infty}^{\infty} (-1)^k \exp [ \ui \ 2 \pi  \ k \ {\cal N}_{\tau^*} (E) ] \ .
\end{equation}
In contrast to the truncated density of states $\rho_{\tau^*}$, this function has the correct delta-peak structure with the same periodic orbit information. Introducing this approximation into the definition of the two--point correlation function we obtain the following expressions

\begin{equation}
r_2(\epsilon_1, \epsilon_2) = \rhob^2(E) + r_2^{(\rm{diag})} (\epsilon_1, \epsilon_2) + r_2^{(\rm{off})} (\epsilon_1, \epsilon_2) \ ,
\end{equation}
with

\begin{eqnarray}
r_2^{(\rm{diag})} (\epsilon_1, \epsilon_2) &=&  \langle \rhot_{\tau^*} (E + \epsilon_1) \rhot_{\tau^*} (E + \epsilon_2) \rangle \label{r2diagA} \ , \\
r_2^{(\rm{off})} (\epsilon_1, \epsilon_2) &=&  \frac{\partial^2}{\partial \epsilon_1 \partial \epsilon_2}
\sum_{k_1,k_2 \neq 0} \frac{ (-1)^{k_1 - k_2} }{4 \pi^2 k_1 k_2} \langle \exp [ \ \ui 2 \pi   k_1  {\cal N}_{\tau^*} (E + \epsilon_1) - \ui 2 \pi   k_2  {\cal N}_{\tau^*} (E + \epsilon_2) ] \rangle \ . \label{r2offA}
\end{eqnarray}
Using the definition (\ref{rhotrunc}), we obtain a double sum over periodic orbits. Expanding the action $S_p (E + \epsilon_j) = S_p(E) + \tau_p \epsilon_j$, and considering that periods up to $\tau^*$ are uncorrelated, it follows that

\begin{equation}
r_2^{(\rm{diag})} (\epsilon_1, \epsilon_2) = \frac{1}{2 \pi^2 \hbar^2} \sum_{\tau_p < \tau^*} \frac{ g_p \tau_p^2 \ \cos ( \tau_p (\epsilon_1 - \epsilon_2) / \hbar ) }{ r_p^2 \left| \det (M_p - I) \right| }  \ .  \label{r2diagB}
\end{equation}
We remark that the sum is made over all periodic orbits (primitive, repetitions, symmetric partners). The factor $g_p$ is usually one, except when the orbit $p$ has a time reversal symmetric partner having the same period and action, in which case $g_p = 2$.

For the off-diagonal term (\ref{r2offA}), after expanding the counting function in the exponential, the average over energy kills all terms with $k_1 \neq k_2$, simplifying to

\begin{equation}
r_2^{(\rm{off})} (\epsilon_1, \epsilon_2) = \frac{1}{4 \pi^2} \frac{\partial^2}{\partial \epsilon_1 \partial \epsilon_2} \sum_{k \neq 0} \frac{ 1 }{k^2} \ue^{ \ui 2 \pi k (\epsilon_1 - \epsilon_2) \rhob } \Phi_k (\epsilon_1, \epsilon_2) \ ,  \label{r2offB}
\end{equation}
with

\begin{equation}
\Phi_k (\epsilon_1, \epsilon_2) = \left\langle \exp \left[ \ \ui 2 \pi   k  [ \Nosc_{\tau^*} (E + \epsilon_1) -  \Nosc_{\tau^*} (E + \epsilon_2) ] \right] \right\rangle \ .  \label{phidef}
\end{equation}
It is known that for chaotic systems, the fluctuating part of the counting function is gaussian distributed \cite{noscgauss}. Thus, we can use the relation $\langle \exp [\ui f(E)] \rangle = \exp [ - \langle f^2 (E) \rangle / 2 ]$, valid for any Gaussian random function $f(E)$ with zero mean. Finally, applying again the diagonal approximation for the correlator of the counting function

\begin{equation}
\Phi_k (\epsilon_1, \epsilon_2) = \left[ \frac{  \Delta_{\tau^*} \left(  (\epsilon_1 - \epsilon_2)/\hbar \right) }{ \Delta_{\tau^*} (0)  }  \right]^{k^2} \ ,  \label{phiC}
\end{equation}
where

\begin{equation}
\Delta_{\tau^*} (u) = \exp \left[ 2 \sum_{\tau_p < \tau^*} \frac{ g_p  \cos (\tau_p u ) }{ r_p^2 \left| \det (M_p - I) \right| } \right] \   \label{Deltadef}
\end{equation}
is a classical function containing all dynamical information of the system. The diagonal term can be also written in terms of this classical function, and noting that both diagonal and off-diagonal terms depend only through the difference $\epsilon = \epsilon_1 - \epsilon_2$, we finally rewrite the two--point correlation function as $r_2 (\epsilon) = \rhob^2 + r_2^{(\rm{diag})} (\epsilon) + r_2^{(\rm{off})} (\epsilon)$ with

\begin{eqnarray}
r_2^{(\rm{diag})} ( \epsilon ) &=&  - \frac{1}{4 \pi^2} \frac{\partial^2}{\partial \epsilon^2} \log \Delta_{\tau^*} \left( \epsilon / \hbar \right) \ , \label{r2diagF} \\
r_2^{(\rm{off})} ( \epsilon ) &=& - \frac{1}{2 \pi^2} \frac{\partial^2}{\partial \epsilon^2} \sum_{k =1}^{\infty} \frac{ \cos ( 2 \pi k \epsilon \rhob ) }{k^2} \left[ \frac{  \Delta_{\tau^*} \left( \epsilon / \hbar \right) }{ \Delta_{\tau^*} (0) } \right]^{k^2} \ . \label{r2offF}
\end{eqnarray}

In order to compare the correlation functions of different systems, or at different energies, to random matrix theory results, it is necessary to unfold the spectrum. The natural local energy scale is the mean level spacing $\delta = 1 / \rhob$, and we define the unfolded two--point correlation function as

\begin{equation}
R_2 (x) = \frac{1}{\rhob^2} r_2 \left( \frac{x}{\rhob} \right) \ ,  \label{R2}
\end{equation}
where $x$ is a dimensionless number. Similarly, $R_2$ can be split into a diagonal and off-diagonal part

\begin{equation}
R_2 ( x ) = 1 + R_2^{(\rm{diag})} ( x ) + R_2^{(\rm{off})} ( x ) \ ,  \label{R2B}
\end{equation}
with

\begin{eqnarray}
R_2^{(\rm{diag})} ( x ) &=&  - \frac{1}{4 \pi^2} \frac{\partial^2}{\partial x^2} \log  \Delta_{\tau^*} \left( x / \hbar \rhob \right) \ , \label{R2diagB} \\
R_2^{(\rm{off})} ( x ) &=& - \frac{1}{2 \pi^2} \frac{\partial^2}{\partial x^2} \sum_{k =1}^{\infty} \frac{ \cos ( 2 \pi k x ) }{k^2} \left[ \frac{  \Delta_{\tau^*} \left( x /  \hbar \rhob \right) }{ \Delta_{\tau^*} (0) } \right]^{k^2} \ . \label{R2offB}
\end{eqnarray}

\subsection{Maps} \label{R2semMaps}

The quantization of classical canonical maps gives a unitary evolution operator $\hat U$ whose eigenvalues $\ue^{\ui \theta_j}$ are on the unit circle. For maps in a compact phase space, the evolution operator is represented by a finite $N$ dimensional unitary matrix, where $N = S / 2 \pi \hbar$, and $S$ is the phase--space area. We define the density of states over the unit circle as

\begin{equation}
\rho (\theta) = \sum_{k=-\infty}^{\infty} \sum_{j=1}^{N} \delta (\theta - \theta_j - 2 \pi k)  \ ,  \label{rhomapsdef}
\end{equation}
where the eigenphases $\theta_j$ are for convenience restricted to the basic period, $ 0 \leq \theta_j < 2 \pi$. By definition $\rho (\theta)$ is a $2 \pi$ periodic function, $\rho (\theta + 2 \pi ) = \rho (\theta)$. As for continuous systems, the density of states can be split into a smooth and fluctuating part $\rho (\theta) = \rhob (\theta) + \rhot (\theta)$. The smooth part is uniform over the unit circle, $\rhob (\theta) = N / 2 \pi$. The fluctuating part can be written as

\begin{equation}
\rhot (\theta) = \frac{1}{2 \pi} \sum_{n=1}^{\infty} [ a_n^* \ue^{\ui n \theta} + a_n \ue^{-\ui n \theta}]  \ ,  \label{rhotildemaps}
\end{equation}
where $a_n = \tr \hat U^{n}$ are the traces of powers of the evolution operator. By unitarity of $\hat U$ we have $a_{-n} =  \tr \hat U^{-n} = a_n^*$. Analogously, a cumulative density of states, or counting function, can be defined as

\begin{equation}
 {\cal N} (\theta) = \int_0^{\theta}  \rho (\theta') {\rm d} \theta' = \Nbar (\theta) + \Nosc (\theta)  \ .\label{Nmaps}
\end{equation}
The smooth part is
\begin{equation}
\Nbar (\theta) = \frac{N}{2 \pi} \theta + C \ ,  \label{Nbarmaps}
\end{equation}
where
\begin{equation}
C = \frac{1}{\pi} \sum_{n=1}^{\infty} \frac{\imag (a_n)}{n} \ ,  \label{Const}
\end{equation}
and the fluctuating part is
\begin{equation}
\Nosc (\theta) = \frac{1}{2 \pi \ui} \sum_{n=1}^{\infty} \left[ \frac{ a_n^* }{n} \ue^{\ui n \theta} - \frac{ a_n }{n} \ue^{-\ui n \theta} \right]  \ .  \label{Noscmaps}
\end{equation}
Both the counting function and its smooth part satisfy the translation relation ${\cal N} (\theta + 2 \pi) = {\cal N} (\theta) + N$, while the fluctuating part is strictly $2 \pi$ periodic.

As for hamiltonian systems, the two--point correlation function is defined as

\begin{equation}
r_2(\epsilon_1, \epsilon_2) =  \langle \rho (\theta + \epsilon_1) \rho (\theta + \epsilon_2) \rangle \ , \label{r2defmaps}
\end{equation}
where the brackets now indicate an average over the phase

\begin{equation}
\langle f (\theta) \rangle = \frac{1}{2 \pi} \int_{0}^{2 \pi} f(\theta') {\rm d} \theta' \ .
\label{averagemaps}
\end{equation}
In terms of the traces of $\hat U^n$, the average for $r_2$ can be computed analytically

\begin{equation}
r_2(\epsilon_1, \epsilon_2) = \left( \frac{N}{2 \pi} \right)^2 + \frac{1}{4 \pi^2} \sum_{n=1}^{\infty} |a_n |^2 [ \ue^{\ui n (\epsilon_1 - \epsilon_2)}  + \ue^{- \ui n (\epsilon_1 - \epsilon_2)}] \ .  \label{r2mapsA}
\end{equation}
It has also been conjectured that for fully chaotic classical maps, in the semiclassical limit, the eigenvalue statistics are the same as for circular ensembles of random matrices. In order to see this semiclassically, it is necessary to estimate the behavior of the traces $a_n$. Tabor \cite{TaborMaps} developed for chaotic maps a trace formula similar to Gutzwiller's,

\begin{equation}
a_n = \tr \hat U^n = \sum_{p^{(n)}} \frac{n \ue^{\ui 2 \pi r_p (N S_p - \nu_p / 4)}}{r_p | \det (M_p - I) |^{1/2} } \ ,  \label{traceformula}
\end{equation}
where the sum is made over all periodic orbits $p$ of period $n$. To compute $|a_n|^2$, necessary for the two--point correlation function, a double sum on periodic orbits is needed. For periods $n \gg 1$ the exponential proliferation of periodic orbits in chaotic maps makes that terms in the double sum coming from different orbits approximately cancel, provided that actions are uncorrelated, and only the diagonal term survives. In the regime $1 \ll n \ll N$, the values of these traces are independent of $N$. For small $n$, the few off-diagonal terms in the double sum on periodic orbits still depend on $N$, but averaging over an ensemble of different $N$ values around a large $N_0$ they will also cancel. It follows that, semiclassically,

\begin{equation}
|a_n|^2 \approx \sum_{p^{(n)}} \frac{n^2 g_p }{r_p^2 | \det (M_p - I) | } \ ,  \label{diagapproxmaps}
\end{equation}
where $g_p = 2$ when the orbit $p$ has a symmetric partner, otherwise $g_p = 1$. By a classical sum rule for chaotic maps, the sum can be further simplified, giving

\begin{equation}
|a_n|^2 \approx g \ n \  .  \label{diagapproxHO}
\end{equation}
However, it is known that for periods $n$ of order $N$, non-trivial correlations among actions of different periodic orbits become important and the diagonal approximation starts to fail. To overcome this problem we adapt the method developed by Bogomolny-Keating to the case of maps, first defining a truncated density of states and counting function

\begin{eqnarray}
\rho_M (\theta) &=&  \frac{N}{2 \pi} + \rhot_M (\theta)  \ ,  \label{} \\
{\cal N}_M (\theta) &=& \frac{N}{2 \pi} \theta + C + \Nosc_M (\theta) \ , \label{}
\end{eqnarray}
with

\begin{eqnarray}
\rhot_M (\theta) &=&  \frac{1}{2 \pi} \sum_{n=1}^{M} [ a_n^* \ue^{\ui n \theta} + a_n \ue^{-\ui n \theta}] \ ,  \label{} \\
\Nosc_M (\theta) &=&  \frac{1}{2 \pi \ui} \sum_{n=1}^{M} \left[ \frac{ a_n^* }{n} \ue^{\ui n \theta} - \frac{ a_n }{n} \ue^{-\ui n \theta} \right] \ . \label{}
\end{eqnarray}
The truncation number $M$ should be of order $N$, in which case the exact eigenphases $\theta_j$ can be well approximated by the condition

\begin{equation}
{\cal N}_M (\theta_j (M) ) = j + 1/2 \ ,  \label{thetaapprox}
\end{equation}
for $ 1 \leq j \leq N$ and $0 \leq \theta_j (M) \leq 2 \pi$. With these approximated eigenphases we define a bootstrapped density of states

\begin{equation}
D_M ( \theta ) =  \sum_{k=-\infty}^{\infty} \sum_{j=1}^{N} \delta (\theta - \theta_j (M) - 2 \pi k) \ .  \label{bootA}
\end{equation}
Using definition (\ref{thetaapprox}) and Poisson summation formula, it can be expressed as

\begin{equation}
D_M ( \theta ) = \rho_M (\theta) \sum_{k=-\infty}^{\infty} (-1)^k \ue^{ \ui 2 \pi {\cal N}_M ( \theta) } \ .  \label{bootC}
\end{equation}
Similarly to Hamiltonian systems, this function has the correct delta--peak structure in contrast to the truncated density of states $\rho_{M}$. Introducing this approximation into the definition of the two--point correlation function we have

\begin{eqnarray}
r_2(\epsilon_1, \epsilon_2) &=&  \langle D_M (\theta + \epsilon_1) D_M (\theta + \epsilon_2) \rangle  \nonumber \\
&=& \langle \rho_M (\theta + \epsilon_1) \rho_M (\theta + \epsilon_2) \sum_{k_1,k_2} (-1)^{k_1 - k_2} \exp [ \ \ui 2 \pi   k_1  {\cal N}_M (\theta + \epsilon_1) - \ui 2 \pi   k_2  {\cal N}_M (\theta + \epsilon_2) ] \rangle \nonumber \\
&=& \frac{N^2}{4 \pi^2} + r_2^{(\rm{diag})} (\epsilon_1, \epsilon_2) + r_2^{(\rm{off})} (\epsilon_1, \epsilon_2) \ .
\end{eqnarray}
In the diagonal part the average on $\theta$ can be done analytically, giving an equation similar to (\ref{r2mapsA}), although the sum is truncated at $M$ and can be rewritten as

\begin{equation}
r_2^{(\rm{diag})} (\epsilon_1, \epsilon_2) = \frac{1}{4 \pi^2} \frac{\partial^2}{\partial \epsilon_1 \partial \epsilon_2} \sum_{n=1}^{M} \frac{|a_n |^2}{n^2} [ \ue^{\ui n (\epsilon_1 - \epsilon_2)}  + \ue^{- \ui n (\epsilon_1 - \epsilon_2)}] \ .  \label{r2mapsdiagA}
\end{equation}
For the off-diagonal term

\begin{equation}
r_2^{(\rm{off})} (\epsilon_1, \epsilon_2) = \frac{1}{4 \pi^2} \frac{\partial^2}{\partial \epsilon_1 \partial \epsilon_2} \sum_{k_1,k_2 \neq 0} \frac{ (-1)^{k_1 - k_2} }{k_1 k_2} \langle \exp [ \ \ui 2 \pi   k_1  {\cal N}_M (\theta + \epsilon_1) - \ui 2 \pi   k_2  {\cal N}_M (\theta + \epsilon_2) ] \rangle \ ,  \label{r2mapsoffA}
\end{equation}
it can be shown that terms with $k_1 \neq k_2$ cancel after averaging over $\theta$, giving

\begin{equation}
r_2^{(\rm{off})} (\epsilon_1, \epsilon_2) = \frac{1}{4 \pi^2} \frac{\partial^2}{\partial \epsilon_1 \partial \epsilon_2}
\sum_{k \neq 0} \frac{ 1 }{k^2} \ue^{\ui N k ( \epsilon_1 - \epsilon_2)} \Phi_k (\epsilon_1, \epsilon_2)  \ ,  \label{r2mapsoffB}
\end{equation}
with

\begin{equation}
\Phi_k (\epsilon_1, \epsilon_2) = \left\langle \exp \left[ \ \ui 2 \pi   k  [ \Nosc_M (\theta + \epsilon_1) -  \Nosc_M (\theta + \epsilon_2) ] \right] \right\rangle \ .  \label{phidefmaps}
\end{equation}
Considering that for chaotic maps the fluctuating part of the counting function is also gaussian distributed, we have

\begin{equation}
\Phi_k (\epsilon_1, \epsilon_2) \approx  \exp \left[ - 2 \pi^2 k^2 \left\langle [ \Nosc_M (\theta + \epsilon_1) -  \Nosc_M (\theta + \epsilon_2) ]^2 \right\rangle \right] = \frac{ \exp \left[ 4 \pi^2 k^2 \left\langle  \Nosc_M (\theta + \epsilon_1)  \Nosc_M (\theta + \epsilon_2) \right\rangle \right] }{ \exp \left[ 4 \pi^2 k^2 \left\langle [ \Nosc_M (\theta) ]^2 \right\rangle \right]  }  \ .  \label{phiBmaps}
\end{equation}
For maps the correlator of the counting function can be computed exactly in terms of the traces $a_n$

\begin{equation}
\left\langle  \Nosc_M (\theta + \epsilon_1)  \Nosc_M (\theta + \epsilon_2) \right\rangle = \frac{1}{4 \pi^2} \sum_{n=1}^{M} \frac{|a_n |^2}{n^2} [ \ue^{\ui n (\epsilon_1 - \epsilon_2)}  + \ue^{- \ui n (\epsilon_1 - \epsilon_2)}] \ .  \label{xxx}
\end{equation}

Introducing the function

\begin{equation}
\Delta_M (\epsilon) = \exp \left[ 2 \sum_{n=1}^{M} \frac{| a_n |^2}{n^2} \cos (n \epsilon)  \right] \ ,  \label{Deltamapsdef}
\end{equation}
both the diagonal and off-diagonal parts of the two--point correlation function are expressed as

\begin{eqnarray}
r_2^{(\rm{diag})} (\epsilon) &=& - \frac{1}{4 \pi^2} \frac{\partial^2}{\partial \epsilon^2} \log \Delta_M (\epsilon) \ ,  \label{r2diagmapsC} \\
r_2^{(\rm{off})} (\epsilon) &=& - \frac{1}{2 \pi^2} \frac{\partial^2}{\partial \epsilon^2} \sum_{k = 1}^{\infty} \frac{1}{k^2} \cos (N k \epsilon) \left( \frac{\Delta_M (\epsilon)}{\Delta_M (0)} \right)^{k^2} \ . \label{r2offmapsC}
\end{eqnarray}

The unfolding for maps is simpler as compared to generic dynamical systems. The smooth density of states is uniform over the unit circle, and one has to divide by a constant. The unfolded variable is $x = \epsilon / \rhob$ giving $\epsilon = 2 \pi x / N$, and the above expressions become

\begin{eqnarray}
R_2^{(\rm{diag})} (x) &=& - \frac{1}{4 \pi^2} \frac{\partial^2}{\partial x^2} \log \Delta_M ( 2 \pi x / N ) \ ,  \label{r2diagmapsD} \\
R_2^{(\rm{off})} (x) &=& - \frac{1}{2 \pi^2} \frac{\partial^2}{\partial x^2} \sum_{k = 1}^{\infty} \frac{1}{k^2} \cos (2 \pi k x) \left( \frac{\Delta_M ( 2 \pi x / N )}{\Delta_M (0)} \right)^{k^2} \ , \label{r2offmapsD}
\end{eqnarray}
where the total unfolded two--point correlation function is $R_2 (x) = 1 + R_2^{(\rm{diag})} (x) + R_2^{(\rm{off})} (x)$. The latter equations show that in the semiclassical limit $N \rightarrow \infty$ the correlation function depends on the behavior of the $\Delta_M$ function at small arguments. We stress that these expressions are completely analogue to those for Hamiltonian systems.

In principle, $\Delta_M$ is a quantum function due to its dependence on the traces $|a_n|^2$. However, as discussed before, this function may be semiclassically approximated by

\begin{equation}
\Delta_M (s) \approx \Delta_M^{\rm (cl)} (s) = \exp \left[ \sum_{n=1}^{M}  \sum_{p^{(n)}} \frac{g_p \cos (n s) }{r_p^2 | \det (M_p - I) |} \right] \ ,  \label{Zclassmaps}
\end{equation}
which is a purely classical function.

\section{Expansion of the classical $\Delta$ function} \label{DeltaFunction}

From the previous section, for Hamiltonian systems we observe that the two--point correlation function is fully determined by the classical function $\Delta_{\tau^*} (u)$. We are interested on the behavior of $R_2 (x)$ for $x$ of order one. Then the argument of the classical $\Delta_{\tau^*}$ function is

\begin{equation}
u = \frac{x}{\hbar \ \rhob } = \frac{2 \pi x}{ \tH } \ ,
\end{equation}
where $\tH$ is the Heisenberg time. In the semiclassical limit this time usually goes to infinity, implying $u \rightarrow 0$. For this reason we are specially interested in the behavior of $\Delta_{\tau^*}$ for small arguments, and we should expand the function around $u=0$. In order to obtain this expansion we write the function in an integral form

\begin{equation}
\log \Delta_{\tau^*} (u) =  2 \int_0^{\tau^*} \frac{1}{\tau^2} K_D (\tau) \cos (u \tau) {\rm d}\tau \ ,  \label{logDeltaA}
\end{equation}
where

\begin{equation}
K_D (\tau) = h^2 \sum_p A_p^2 \ \delta (\tau -  \tau_p) =  \sum_p \frac{ g_p \ \tau_p^2 \ \delta (\tau -  \tau_p) }{ r_p^2 \left| \det (M_p - I) \right| }  \ ,  \label{KDdef}
\end{equation}
is the diagonal form factor. For chaotic systems this sum may be estimated by Hannay-Ozorio de Almeida sum rule,

\begin{equation}
K_D (\tau) \approx g \tau  \ .  \label{KDapprox}
\end{equation}
This approximation is valid for $\tau \gg \tmin$, where $\tmin$ is the period of the shortest periodic orbit. The $g$ factor is 2 for systems with time-reversal symmetry, otherwise $g=1$. For sufficiently large $\tau^*$, we can consider an arbitrary intermediate time $\tau_1$, such that $\tmin \ll \tau_1 \ll \tau^*$, then

\begin{equation}
\log \Delta_{\tau^*} (u) = 2 \sum_{\tau_p < \tau_1} \frac{ g_p  \ \cos ( u \tau_p) }{ r_p^2 \left| \det (M_p - I) \right| } + 2 g \int_{\tau_1}^{\tau^*} \frac{ \cos (u \tau) }{\tau} {\rm d}\tau \ .  \label{logDeltaB}
\end{equation}
The last integral can be solved in terms of the cosine integral function

\begin{equation}
{\rm Ci} (x) = - \int_x^{\infty} \frac{\cos u}{u} {\rm d}u = \log |x| + \gamma + \sum_{l=1}^{\infty} \frac{(-1)^l x^{2 l}}{2 l (2 l)!} \ ,  \label{CosInt}
\end{equation}
where $\gamma = 0.5772 \ldots$ is the Euler-Mascheroni constant, giving

\begin{equation}
\log \Delta_{\tau^*} (u) = 2 \sum_{\tau_p < \tau_1} \frac{ g_p  \ \cos ( u \tau_p) }{ r_p^2 \left| \det (M_p - I) \right| } + 2 g {\rm Ci} (u \tau^*) - 2 g {\rm Ci} (u \tau_1) \ .  \label{logDeltaC}
\end{equation}

Expanding the cosine into the periodic orbit sum for small $u$, and using the Taylor series (\ref{CosInt}) for the cosine integral function, we obtain

\begin{equation}
\log \Delta_{\tau^*} (u) =  2 g \left[ d_0 + \sum_{l=1}^{\infty} d_l u^{2 l} - \log u - \gamma + {\rm Ci} (u \tau^*) \right] \ , \label{logDeltaD}
\end{equation}
with

\begin{eqnarray}
d_0 &=& \lim_{\tau_1 \rightarrow \infty} \left[ \sum_{\tau_p < \tau_1} \frac{ g_p / g}{ r_p^2 \left| \det (M_p - I) \right| } - \log \tau_1 \right]  \ ,  \label{d0} \\
d_l &=& \frac{(-1)^l}{(2 l)! } \lim_{\tau_1 \rightarrow \infty} \left[ \sum_{\tau_p < \tau_1} \frac{ (g_p / g) \tau_p^{2 l} }{ r_p^2 \left| \det (M_p - I) \right| } - \frac{ \tau_1^{2 l} }{ 2 l } \right] \ . \label{dl}
\end{eqnarray}
The latter constants depend on classical information. To compute $\Delta_{\tau^*} (0)$, necessary for the off-diagonal part of $R_2$, we easily solve the integral of Eq. (\ref{logDeltaB}) arriving to

\begin{equation}
\log \Delta_{\tau^*} (0) = 2 g [ d_0 + \log \tau^* ] \ .  \label{logDelta0}
\end{equation}


For maps, in the semiclassical limit $N \rightarrow \infty$, provided that there exist a $n^*$ such $1 \ll n^* \ll M$, the sum over traces in Eq. (\ref{Deltamapsdef}) can be split

\begin{equation}
\log \Delta_M (\epsilon) =  2 \sum_{n=1}^{n^*} \cos (n \epsilon) \frac{|a_n |^2}{n^2} + 2 g \sum_{n=n^*+1}^{M} \frac{\cos (n \epsilon)}{n}  \ ,  \label{DeltamapsA}
\end{equation}
where in the second sum we have used the classical summation rule (\ref{diagapproxHO}), $|a_n |^2 \approx g n$. After rearranging terms we have

\begin{equation}
\log \Delta_M (\epsilon) = 2 g \sum_{n=1}^{n^*} b_n \cos (n \epsilon) + 2 g \sum_{n=1}^{M} \frac{\cos (n \epsilon)}{n}  \ .  \label{DeltamapsB}
\end{equation}
with
\begin{equation}
b_n = \frac{|a_n |^2}{g n^2} - \frac{1}{n} \approx \sum_{p^{(n)}} \frac{ (g_p / g) }{r_p^2 | \det (M_p - I) |} - \frac{1}{n} \ .  \label{bndef}
\end{equation}
These coefficients measure the deviation of traces with respect to the classical sum rule. Therefore it is expected that $b_n \rightarrow 0$ as $n \rightarrow \infty$. Moreover, if $b_n$ decreases sufficiently fast as $n$ increases, the first sum becomes independent of $n^*$, and we can expand the cosine for small $\epsilon$. The first sum in (\ref{DeltamapsB}) is

\begin{equation}
\sum_{n=1}^{n^*} b_n \cos (n \epsilon) = \sum_{l=0}^{\infty} c_l \epsilon^{2 l}  \ ,  \label{firstsum}
\end{equation}
with

\begin{equation}
c_l = \frac{(-1)^l}{(2 l)!} \lim_{n^* \rightarrow \infty} \sum_{n=1}^{n^*} b_n n^{2 l} \ .  \label{dlmaps}
\end{equation}
If $M \rightarrow \infty$, the second sum in (\ref{DeltamapsB}) has an analytic solution. The error due to a finite $M$ can be estimated through the Euler-Maclaurin formula

\begin{equation}
\sum_{n=1}^{M} \frac{\cos (n \epsilon)}{n} = - \log | 2 \sin (\epsilon / 2) | + F_M (\epsilon) \ ,  \label{SumCos}
\end{equation}
with

\begin{equation}
F_M (\epsilon) = - \sum_{n=M + 1}^{\infty} \frac{\cos (n \epsilon)}{n} = \Ci (M \epsilon) + {\cal O} (M^{-1}) \ .  \label{FMdef}
\end{equation}
For fixed $\epsilon$, as $M \rightarrow \infty$, $F_M (\epsilon)$ is small. Finally

\begin{equation}
\Delta_M (\epsilon) = \frac{1}{| 2 \sin (\epsilon / 2) |^{2g} } \exp \left\{ 2 g \left[ c_0 + \sum_{l=1}^{\infty} c_l \epsilon^{2 l} + F_M (\epsilon) \right] \right\}  \ .  \label{DeltamapsC}
\end{equation}

For $\epsilon = 0$, we evaluate Eq. (\ref{DeltamapsB}) directly
\begin{equation}
\log \Delta_M (0) = 2 g \sum_{n=1}^{n^*} b_n + 2 g \sum_{n=1}^{M} \frac{1}{n} = 2 g [ c_0 + H_M ] \ ,  \label{Deltamaps0}
\end{equation}
where $H_M$ is the harmonic number. It is known that for large $M$ the harmonic number admits the expansion

\begin{equation}
H_M = \log M + \gamma + \frac{1}{2 M} - \frac{1}{12 M^2} + {\cal O} ( M^{-4} )  \ ,  \label{HM}
\end{equation}
and finally

\begin{equation}
\Delta_M (0) = M^{2g} \exp \left\{ 2 g \left[ c_0 + \gamma + \frac{1}{2 M} - \frac{1}{12 M^2} + {\cal O} ( M^{-4} ) \right] \right\}  \ .  \label{Deltamaps0B}
\end{equation}

\section{Comparison to random matrix ensembles} \label{R2RMT}

In RMT, correlation functions and other related statistics are computed averaging over the ensemble, instead of averaging over the spectrum for a particular realization of the ensemble (although it is known that for sufficiently large matrices both results coincide). In several cases, exact analytical expressions are obtained. We are interested specially in random matrix ensembles of finite dimension. For the CUE of dimension $N$, a simple exact expression for the two--point correlation function is \cite{Mehta}

\begin{equation}
R_2 (x) = 1 - \left[ \frac{ \sin (\pi x ) }{ N \sin ( \pi x / N ) }\right]^2 \ .  \label{R2CUEN}
\end{equation}
As $N \rightarrow \infty$ it expands as

\begin{equation}
R_2 (x) = 1 - \left[ \frac{ \sin (\pi x ) }{ \pi x }\right]^2 - \frac{1}{3 N^2} \sin^2 (\pi x) - \frac{1}{15 N^4} [ \pi x \sin (\pi x) ]^2 + {\cal O} (N^{-6}) \ ,  \label{R2CUEapprox}
\end{equation}
in terms of even powers of the dimension $N$. The leading order is the usual RMT result for the $R_2$ function, that also coincides with that of the Gaussian Unitary Ensemble (GUE) in the limit $N \rightarrow \infty$.

\subsection{Hamiltonian systems} \label{compHamiltonian}

Inserting Eqs. (\ref{logDeltaD}) and (\ref{logDelta0}) for $\Delta_{\tau^*}$ into Eqs. (\ref{R2diagB}) and (\ref{R2offB}) we can compute the unfolded two--point correlation function in terms of classical quantities $d_l$. In order to simplify the equations and compare them to random matrix results we have to make further approximations:

i) For small but fixed $u$, if we consider $\tau^* \rightarrow \infty$, the term $\Ci (u \tau^*)$ in Eq. (\ref{logDeltaD}) vanishes.

ii) As we considered $\tau^* \sim \tH$, and $\Delta_{\tau^*} (0) \sim (\tau^*)^{2 g}$, we keep only the term $k = 1$ in the sum of the off-diagonal part.

iii) For the second derivative in the off-diagonal part we neglect the derivatives of $\Delta(u)_{\tau^*}$ because they give higher order terms in $\tH$.

The two--point correlation function for dynamical systems, up to second order in Heisenberg time, simplifies to

\begin{eqnarray}
R_2^{(\rm{diag})} ( x ) &=&  - \frac{g}{2 \pi^2} \left[ \frac{1}{x^2} + 2 d_1 \left( \frac{2 \pi}{\tH} \right)^{2} + {\cal O} (\tH^{-4}) \right] \ , \label{R2diagD} \\
R_2^{(\rm{off})} ( x ) &=& 2 \left( \frac{\tH}{2 \pi \ue^{\gamma} \tau^* x } \right)^{2 g} [ 1 - 2 \sin^2 (\pi x ) ] \exp \left[ 2 g d_1 (2 \pi x / \tH )^{2} + {\cal O} (\tH^{-4}) \right] \ ,  \label{R2offD}
\end{eqnarray}
Moreover, following the same argument as in Ref.\cite{BogKea96}, for systems without time-reversal symmetry ($g=1$) we can take $\tau^* = \tH / \ue^{\gamma}$, in order to exactly cancel the divergence in the diagonal part by the off-diagonal part as $x \rightarrow 0$, arriving to

\begin{equation}
R_2 ( x ) = 1 - \left[ \frac{ \sin (\pi x ) }{ \pi x }\right]^2 - \frac{8 d_1}{ \tH ^2 } \sin^2 (\pi x) + {\cal O} (\tH^{-4}) \ .  \label{R2g1final}
\end{equation}
We see explicitly that in the limit $\tH \rightarrow \infty$ we recover the asymptotic CUE or GUE result, and we have a first order correction for finite Heisenberg time depending on periodic orbits through the coefficient $d_1$. Moreover, comparing to Eq. (\ref{R2CUEapprox}) for the CUE ensemble, we see that the functional form in $x$ of the first correction is the same. Equating the constants it follows that a generic chaotic dynamical system without time-reversal symmetry has the same two point function as a finite CUE ensemble of dimension

\begin{equation}
N_{\rm eff} = \frac{ \tH }{ \sqrt{24 d_1} } \ ,
\end{equation}
where the value of $d_1$ was given in Eq. (\ref{d1}).

An estimate of the characteristic time associated to $d_1$ may be obtained as follows. The sum over periodic orbits in Eq.(\ref{d1}) may be approximated by
\begin{equation}
\int_{\tmin}^{\infty} d \tau \chi (\tau) \frac{\tau^2}{ \left| \det (M_{\tau} - I) \right|} \label{int}
\end{equation}
where $\chi(\tau) = \exp (\htop \tau)/\tau - \exp(\htop \tau/2)/2\tau - \ldots$ is the smooth density of primitive periodic orbits of period $\tau$, and $\htop$ is the topological entropy (cf Ref.\cite{leb2004}). Using the approximate sum rule
$\left| \det (M_{\tau} - I) \right| \approx \exp(\htop \tau)$, it follows from Eq.(\ref{int}) and the definition of $d_1$ that
\begin{equation}
d_1 \approx \frac{\tmin^2}{4} + \frac{1}{\htop^2} \left( \frac{\htop \tmin}{2} + 1 \right) \ue^{- \htop \tmin /2}
\end{equation}

\subsection{Maps} \label{compMaps}

We compute the unfolded two--point correlation function for maps inserting Eqs. (\ref{DeltamapsC}) and (\ref{Deltamaps0B}) for $\Delta_M$ into Eqs. (\ref{r2diagmapsD}) and (\ref{r2offmapsD}),

\begin{eqnarray}
R_2^{(\rm{diag})} ( x ) &=&  - \frac{g}{2 \pi^2} \left[ \frac{ (\pi / N)^2 }{ \sin^2 (\pi x / N)} + 2 c_1 \left( \frac{ 2 \pi }{N} \right)^{2} + {\cal O} (N^{-4}) \right] \ , \label{R2diagmapF} \\
R_2^{(\rm{off})} ( x ) &=& \frac{2}{ [ 2 M \ue^{\gamma} \sin (\pi x / N) ]^{2 g} } [ 1 - 2 \sin^2 (\pi x ) ] \exp \left[ 2 g c_1 (2 \pi x / N)^{2}  + {\cal O} (N^{-4}) \right] \ ,  \label{R2offmapF}
\end{eqnarray}
All the information about the map is contained in the constant $c_1$, that can be computed semiclassically through the traces $a_n$. For maps without time-reversal symmetry ($g=1$), if we take $M = N / \ue^{\gamma}$, and if we first consider the constant $c_1 = 0$, we arrive to

\begin{equation}
R_2 (x) = 1 - \left[ \frac{ \sin (\pi x ) }{ N \sin ( \pi x / N ) }\right]^2 \ ,  \label{R2mapsF}
\end{equation}
which is exactly the same exact result (\ref{R2CUEN}) for $N$-dimensional CUE matrices. This would show that evolution operator matrices of chaotic maps without time-reversal symmetry behave as CUE random matrices of the same dimension.

Nevertheless, considering that constant $c_1$ is finite, up to second order in $N$, the two--point correlation function is

\begin{equation}
R_2 ( x ) = 1 - \left[ \frac{ \sin (\pi x ) }{ \pi x }\right]^2 - \frac{ (1 + 24 c_1) }{3 N^2} \sin^2 (\pi x) + {\cal O} (N^{-4}) \ .  \label{R2mapsg1final}
\end{equation}
Comparing to Eq. (\ref{R2CUEapprox}), we conclude that corrections to the asymptotic result are the same as a CUE matrix of effective dimension

\begin{equation}
N_{\rm eff} = \frac{ N }{ \sqrt{1 + 24 c_1} } \ .
\end{equation}
The value of constant $c_1$ for maps was explicitly given in Eq. (\ref{d1maps}). Similarly to Hamiltonian systems, an estimation for $c_1$ can be done from the density of periodic orbits. Here the integral over time is replaced by a sum over integer periods $n$, giving

\begin{equation}
c_1 \approx \frac{ \ue^{- \htop / 2}  }{ 4 (1 - \ue^{- \htop / 2} )^2 } \ .
\end{equation}
where now the topological entropy $\htop$ is a dimensionless number.

\section{Conclusions} \label{Conclusions}

The results obtained here for the two--point correlation function may, hopefully, be extended to arbitrary $n$-point functions (see the analoguous discussion for the Riemann Zeta function in Refs. \cite{BBLM} and \cite{Bogo2007}). This extension would imply that all statistical properties of eigenvalues of chaotic systems without time reversal symmetry will also agree with that of circular random matrices of effective dimension $N_{\rm eff} = \tH / \sqrt{24 d_1}$.  In particular, this should be the case for the nearest neighbor spacing distribution.

The example of the Riemann Zeta function allows to make explicit comparisons with our results, due to the huge numerical data on zeroes and explicit periodic orbit information \cite{odlyzko}. Following \cite{BerryRiemann}, the prime numbers $p$ label the primitive periodic orbits of period $\tau_p = \log p$, with $\left| \det (M_p - I) \right| = p^r$. With these replacements, we get

\begin{equation}
d_1 = \frac{1}{2} \lim_{\tau_1 \rightarrow \infty} \left[ \frac{ \tau_1^2 }{2} - \sum_{r \log p < \tau_1} \frac{ \log ^2 p }{ p^r }  \right] \ . \label{d1Riemann}
\end{equation}
where the sum is made over primes $p$ and integers $r$. This limit gives $d_1 = 0.78657$, which is exactly half the value of $\beta = 1.57314 \ldots$ obtained in \cite{BBLM}, and gives in Eq.(\ref{neff}) the correct effective dimension. In Ref.\cite{BBLM}, the nearest neighbor spacing distribution was numerically checked for different heights $E$ in the spectrum with high precision.

For general chaotic dynamical systems, the computation of $d_1$ needs explicit classical information on periodic orbits. However, for $\tau_1 \gg \tmin$, the sum will converge towards $\tau_1^2/2$. Therefore, the numerical computation of $d_1$ relies mainly on the computation of the contribution of short periodic orbits.

For maps, the constant $c_1$ can be computed using the quantum values of the traces or, alternatively, their semiclassical approximations in terms of periodic points.

One may wonder if GUE asymptotics may be used instead of CUE. Our analysis indicate that the correct ensemble is the latter since, due to a non uniform density of eigenvalues, the GUE has a different asymptotics compare to Eq.(\ref{R2CUEapprox}). In particular, the first correction goes like $(-1)^N/N$, which is not derived from our semiclassical expansion.

It would be also desirable to extend this study to systems with time--reversal symmetry, and compare to the Circular Orthogonal Ensemble for finite $N$, although expressions are more complex even in the limiting case $N \rightarrow \infty$.

\vspace{0.5cm}
\noindent
{\bf Acknowledgements}
\vspace{0.25cm}

\noindent The authors wants to warmly thank Oriol Bohigas and Eug\`ene Bogomolny for useful discussions.

\end{document}